\documentclass[aps,pra,reprint]{revtex4-1}
\usepackage[T2A]{fontenc}
\usepackage[utf8]{inputenc}
\usepackage[russian,english]{babel}
\usepackage{graphicx}
\usepackage{amsmath}
\usepackage{amssymb}

\usepackage{comment}

\begin{document}

\title{Asymptotic electron motion in strong radiation-dominated regime}

\date{\today}
\author{A.~S.~Samsonov}
\author{E.~N.~Nerush}
\email{nerush@appl.sci-nnov.ru}
\author{I.~Yu.~Kostyukov}
\affiliation{Institute of Applied Physics of the Russian Academy of
Sciences, 46 Ulyanov St., Nizhny Novgorod 603950, Russia}

\begin{abstract}
    We study electron motion in electromagnetic (EM) fields in the radiation-dominated regime.
    It is shown that the electron trajectories become close to some asymptotic trajectories in the
    strong field limit. The description of the electron dynamics by this
    asymptotic trajectories significantly differs from the ponderomotive description that is barely
    applicable in the radiation-dominated regime. The particle velocity on the asymptotic
    trajectory is completely determined by the local and instant EM field. The general properties
    of the asymptotic trajectories are discussed. In most of standing EM waves (including identical
    tightly-focused counter-propagating beams) the asymptotic trajectories are periodic with the
    period of the wave field. Furthermore, for a certain model of the laser beam we show that the
    asymptotic trajectories are periodic in the reference frame moving along  the beam with its group
    velocity that may explain the effect of the radiation-reaction trapping.
\end{abstract}

\maketitle

\section{Introduction}

If the amplitude of an optical field is such that an electron gains in it energy of hundreds of
its
rest-mass energy, the electron starts to emit synchrotron radiation and can lose its energy
efficiently~\cite{Bulanov04}. This phenomenon --- radiation reaction --- is highly important for theoretical physics
and astrophysics, therefore the motion of electrons in strong laser field nowadays is a topic of
numerous theoretical investigations~\cite{Di09, Duclous11, Nerush11b, Thomas12, Neitz13, Vranic16},
and it has been studied recently in the experiments~\cite{Cole17, Poder17}.  Also, the emission of
hard photons by electrons in a strong laser field lets one to make a femtosecond broadband source
of MeV photons, based on either laser pulse -- electron beam collision~\cite{Corde13, Sarri14,
Yan17}, laser-plasma interaction~\cite{Ridgers12, Bashinov13, Nerush14, JI14a, Li17} or
electromagnetic cascades~\cite{Nerush11a, Gonoskov17b}.

In the interaction of a strong laser pulse with a plasma, radiation losses can significantly
affect the plasma dynamics, and, for instance, lead to less-efficient ion
acceleration~\cite{Tamburini10, Tamburini12, Capdessus12, Capdessus15, Nerush15}, the enhancement of
the laser-driven plasma wakefield~\cite{Gelfer18a, Gelfer18b}, highly efficient
laser pulse absorption~\cite{Grismayer16}, relativistic transparency reduction~\cite{Zhang15}, and
to the inverse Faraday effect~\cite{Liseykina16}.

Despite of high importance of the radiation losses
for laser-plasma physics at high intensity, there is no general concept of the losses impact
on the electron motion, and this impact is considered mostly by \textit{ad hoc}
hypotheses and particle-in-cell (PIC) simulations.  Only for a few field configurations the
analytical solutions for motion of emitting electron are present~\cite{Zeldovich75, Di08a,
Kostyukov16a}, whereas for the motion of the non-emitting electron the Miller's ponderomotive
concept~\cite{Miller58} is applicable in a vast number of cases.

\begin{figure}
	\includegraphics[width=1\linewidth]{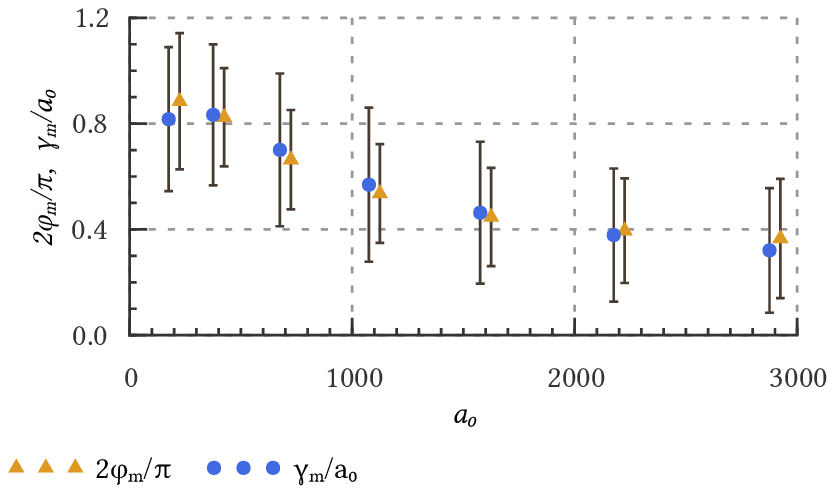}
    \caption{\label{circe}Electrons moving in the rotating electric field and experiencing quantum
    radiation reaction for many periods of the rotation: (circles) the ratio of the mean Lorentz factor
    $\gamma_m$ to $a_0$ and  (triangles) the
    mean angle $\varphi_m$ between the particle velocity and the vector opposite to the electric field, for different
    values of the field amplitude $a_0$. Bars depict the standard deviations $\pm \sigma$. Results
    of PIC-MC simulations for the field angular frequency $\omega = 2 \pi c / \lambda$, $\lambda
    = 1 \text{ } \mu \text{m}$.}
\end{figure}

In the high-intensity field, the energy gained by the electron can be
significantly limited by the radiation losses. In this case, in contrast to the low-intensity
limit, the electron Lorentz factor becomes small in comparison with the field amplitude: $\gamma /
a_0 \ll 1$; here
$\gamma$ is the electron Lorentz factor and $a_0 = e E_0 / mc \omega$ is the normalized amplitude
of the electric field, $E_0$, $\omega$ is the typical angular frequency of the field, $c$ is the speed of
light, $m$ and $e>0$ are the electron mass and the magnitude of the electron charge, respectively.
The smallness of $\gamma / a_0$ allows one to simplify the analytical treatment of the electron motion
in the strong radiation-dominated regime. This can be illustrated by a stationary Zel'dowich's
solution~\cite{Zeldovich75} for the electron motion in the rotating electric field $\mathbf E(t)$. At
moderate field intensity the angle $\varphi$ (between the particle velocity and the vector
$-\mathbf E$) is
connected with the electron Lorentz factor $\gamma$. However, in the strong radiation dominated
regime $\varphi$ and $\gamma / a_0$ tend to zero (see Fig.~\ref{circe}), and the particle velocity
coincides with the direction
of the electric field ($\mathbf{v \parallel E}$), thus $\gamma$ is not needed in order to compute the particle trajectory.

In Refs.~\cite{Fedotov14b, Gonoskov17} the concept of the electron motion, that in the
radiation-dominated regime can supersede the ponderomotive concept, is discussed. It have been shown
in Ref.~\cite{Fedotov14b} that in the regime of dominated radiation friction the number of degrees
of freedom, which govern the electron motion, is reduced. Namely, it is shown for the rotating
electric field with the Gaussian envelope, that on the time scales larger than the rotation period,
the electron position is described by a first-order differential equation that does not contain the
electron momentum. For this, the electron motion with Landau--Lifshitz radiation reaction have been
considered. It is also shown that in the radiation-dominated regime,
electrons are not expelled from but are captured for a long time by the strong-field region.

In Ref.~\cite{Gonoskov17} it is shown for almost arbitrary field configuration, that in the strong
field limit in a timescale, much smaller than the timescale of the field variation, the direction of the
electron velocity approaches some certain direction that is determined only by the values of the
local electric and magnetic fields. Then, as the electron velocity is known, the electron
trajectory can be reconstructed. This approach, called the ``low-energy limit'', was used (but not
described) for the fields of the linearly polarized standing waves earlier~\cite{Gonoskov14}.

Let us emphasize that if the electron velocity is determined not by the electron momentum but
by the local fields, one can describe the plasma dynamics with hydrodynamical equations. Indeed,
in this case the currents in the Maxwell's equations depend only on the particle density and
particle velocity (i.e. on the particle density and local EM fields), therefore the first-order
equation for the electron position together with the
Maxwell's equations and the continuity equation form a closed system of equations.

In this paper we present the first step toward such a hydrodynamical approach to the plasma dynamics in
the radiation-dominated regime. Namely, in Sec.~\ref{estimates} we estimate $\gamma / a_0$ ratio
and the threshold of the radiation-dominated regime. In Sec.~\ref{Asymptote}
for arbitrary field configuration we find the first-order equation for the electron position, by a
method different from Ref.~\cite{Gonoskov17} and with B-case (see below)
considered separately. The right-hand-side of this equation is the velocity field that is fully
determined by the local field vectors.
It is shown that $\gamma \ll a_0$ is enough for this first-order equation to be valid in the laser
field.
In Sec.~\ref{tests} we compare the solution of this first-order equation
with the solution of the exact equations of the electron motion for a number of field configurations.
In Sec.~\ref{absorption-induced-trapping} we discuss the relation between the velocity field and
the Poynting vector. In
Sec.~\ref{properties} the symmetry of the velocity field induced by the
symmetry of the Maxwell's equations, is considered, and the dramatic difference between the
ponderomotive description and the description by the velocity field in the radiation-dominated
regime is demonstrated. Thus, in the subsection~\ref{standing-waves}, in the limit of strong fields, the electron motion
in a wide class of periodic standing waves is shown to be periodic. From this, in the
subsection~\ref{TE11} we show with a certain model of the laser beam that the beam
can capture the electrons and carries them along itself with the beam group velocity.
Sec.~\ref{conclusion} is the conclusion.

\section{Strong radiation-dominated regime}
\label{estimates}

In order to estimate the threshold value of the normalized field amplitude $a_0$ for the
radiation-dominated regime, let us start from
the equations of the electron motion with the Landau--Lifshitz radiation reaction force
incorporated:
\begin{eqnarray}
    \label{dpdt}
    \frac{d\mathbf p}{dt} = -\mathbf E - \mathbf v \times \mathbf B - F_{rr} \mathbf v, \\
	\label{dgdt}
    \frac{d\gamma}{dt} = -\mathbf v \mathbf E - F_{rr} v^2,
\end{eqnarray}
where time is normalized to $1/\omega$, $\mathbf v$ is the electron velocity normalized to the speed of light $c$, $\mathbf p =
\gamma \mathbf v$ is the electron momentum normalized to $mc$, $\mathbf E$ and $\mathbf B$ are
the electric and the magnetic fields respectively (normalized to $mc\omega/e$),
and $F_{rr} \mathbf v$ is the main term of the radiation reaction force~\cite{LandauII}:
\begin{equation}
    \label{Frr}
    F_{rr} = \alpha \gamma^2 \frac{2}{3} \frac{\hbar \omega}{mc^2} \left\{(\mathbf{E+v \times B})^2
    - (\mathbf{Ev})^2 \right\}.
\end{equation}
Here $\alpha = e^2 / \hbar c \approx 1 / 137$ is the fine-structure constant and $\omega$ is the frequency characterizing the
time-scale or the space-scale of the field (e.g. angular frequency of the laser field).

The radiation losses increase sharply with the increase of $\gamma$, therefore for some electron Lorentz
factor $\gamma = \bar \gamma$, the further energy gain stops due to the losses. The corresponding
value $\bar \gamma$ can be found from Eq.~\eqref{dgdt} assuming that the transverse and the longitudinal
to $\mathbf v$ components of the Lorentz force are of the order of $a_0$:
\begin{equation}
    \bar \gamma \approx \sqrt{\frac{3}{2 \alpha a_0} \frac{mc^2}{\hbar \omega}},
\end{equation}
where $a_0$ is the characteristic electric field strength.
In the absence of the radiation reaction the electron energy in the field can be estimated as
$\gamma \sim a_0$, thus the radiation-dominated regime corresponds to $\bar \gamma \ll a_0$ hence
\begin{equation}
    a_0 \gg a_0^* = \left(\frac{3}{2 \alpha} \frac{mc^2}{\hbar \omega} \right)^{1/3}.
\end{equation}
Note for the laser wavelength $\lambda = 1 \text{ }\mu\text{m}$ the amplitude $a_0^* \approx 440$ that
corresponds to the intensity $I \approx 5 \times 10^{23} \text{ W}\, \text{cm}^{-2}$. This level of intensity is
expected to be reached in the near future with such facilities as
ELI-beamlines~\cite{Garrec14}, ELI-NP~\cite{ELI-NP}, Apollon~\cite{Apollon}, Vulcan 2020~\cite{Vulcan2020}, or
XCELS~\cite{Bashinov14}.

In the case of strong radiation losses the angle between the Lorentz force and the
electron velocity can be small, and the transverse to $\mathbf v$ component of the Lorentz force
becomes much lower than the longitudinal one. However, this doesn't affect much the given
estimates. For instance, for the electron motion in the rotating electric field from the stationary
Zel'dowich's solution~\cite{Zeldovich75} we get $\varphi
\approx \gamma / a_0$ and $\gamma \approx (a_0 / \mu)^{1/4} \ll a_0$ (where $\mu = 2 \alpha \hbar \omega / 3
mc^2$) at $a_0 \gg a_0^*$, with the same estimate for $a_0^*$ (except the factor $3/2$ in the
parentheses, see Ref.~\cite{Zeldovich75}). Note also that the quantum consideration of the radiation reaction gives results that are
close to the Zel'dovich's solution: in the Monte Carlo (MC) simulations the mean $\varphi$ value is about $\pi
/ 2$ times larger than $\gamma / a_0$, and $\gamma / a_0$ drops with the increase of the field
amplitude (Fig.~\ref{circe}).

In what follows we assume that the field is far beyond the threshold of the radiation-dominated
regime, $a_0 \gg a_0^*$.

\section{Velocity field and asymptotic trajectories}
\label{Asymptote}

The reduced equations of the electron motion for arbitrary field configuration can be derived as
follows. The equation for the electron velocity can be obtained from Eqs.~\eqref{dgdt} and
\eqref{dpdt}, and is the following:
\begin{equation}
    \label{dvdt}
    \frac{d\mathbf v}{dt} = -\frac{1}{\gamma}\left\{ \mathbf{E - v (v E) + v \times  B} +
    \frac{F_{rr} \mathbf v}{\gamma^2} \right\},
\end{equation}
where the first three terms in the parentheses approximately correspond to the transverse to $\mathbf v$
component of the Lorentz force.

If the angle $\psi$ between the Lorentz force and the electron velocity
is noticeable
($\psi \sim 1$), then the term with
$F_{rr}$ in Eq.~\eqref{dvdt} is negligible, because $F_{rr} / \gamma^2 \sim a_0^2 \alpha \hbar \omega /
mc^2 \ll a_0$ for reasonable field amplitudes, $E_0 \lesssim E_S / \alpha$, where $E_S = m^2 c^3 / e
\hbar$ is the Sauter--Schwinger critical field. Thus, as far as
$\gamma \ll a_0$, we have $|d \mathbf v / dt|
\gg 1$.

It means that the characteristic timescale of the velocity vector variation is small,
$\tau_{\mathbf v} \sim \gamma / a_0 \ll 1$. Therefore, on small time scales it can be assumed that the fields
$\mathbf E$ and $\mathbf B$ in Eq.~\eqref{dvdt} are constant. In the constant EM field the
electron velocity $\mathbf v$ in a time of some $\tau_{\mathbf v}$ approaches some asymptotic
direction. This direction
corresponds to $\psi \to 0$ hence $d \mathbf v / dt = 0$, and can be found as follows.

\subsection{B-case}

In the case $\mathbf{E \cdot B} = 0$ and $B > E$ there is a reference frame $K'$ in which the field
is
purely magnetic, and $\mathbf{B}' \parallel \mathbf{B}$ (here strokes denote quantities in $K'$).
In $K'$ the electron goes along the helical path with its axis parallel to the
direction of $\mathbf B'$. The corresponding drift velocity of the electron in the laboratory
reference frame $K$ is the speed of $K'$ in $K$ and can be found from the following equation:
\begin{equation}
\label{eta}
\mathbf{E + v \times B} = 0.
\end{equation}
Let us note that Eq.~\eqref{eta} does not depend on the component of the velocity parallel to the
magnetic field,
so one can choose this component arbitrarily (implying $v < 1$). One can choose, for example,
the solution with $\mathbf{v \cdot B} = 0$, i.e.:
\begin{equation}
\label{eta_explicit}
    \mathbf v = \frac{\mathbf{E \times B}}{B^2}.
\end{equation}
As shown in Sec.~\ref{TE11} the ambiguity of $\mathbf v$ in this case can be resolved by
additional physical considerations.

\subsection{E-case}

If $E \cdot B \neq 0$ or $E > B$ there is a reference frame $K'$, in which $\mathbf{E'} \parallel
\mathbf{B'}$ or $B' = 0$. The electron trajectory in $K'$ asymptotically approaches the straight
line parallel to $\mathbf{E}'$, and $v$ approaches $1$. Note that for the resulting electron trajectory $\mathbf{ v
\cdot E} < 0$ as far as the electron is accelerating by the field.

As $v \approx 1$ and the electron moves along the straight line, in the laboratory reference frame
$K$ the resulting $\mathbf v$ can be found from the equation $d\mathbf{v}/dt = 0$, that yields
\begin{equation}
\label{beta}
\mathbf{E - v (v E) + v \times  B} = 0,
\end{equation}
Scalar multiplication of Eq. (\ref{beta}) by
$\mathbf B$, $\mathbf E$ and $\mathbf{E \times B}$ leads to the following solution:
\begin{eqnarray}
    \label{betaB}
    \mathbf{vB = \frac{EB}{vE}}, \\
    \label{betaEB}
    \mathbf{v \cdot E \times B} = E^2 - (\mathbf{v E})^2, \\
    \label{betaE}
    \mathbf{vE} = -\sqrt{\frac{E^2 - B^2 + \sqrt{(E^2-B^2)^2 +
    4\mathbf{(EB)^2}}}{2}}, \\
    \label{betaEEB}
    \mathbf{v \cdot E \times [E \times B]} = \mathbf{(v E) (E B) -
    (v B)} E^2.
\end{eqnarray}
The right-hand-side of Eq.~\eqref{betaE} is relativistic invariant, and we choose the sign ``$-$'' in
order to obtain the stable trajectory in $K'$. For the opposite sign, ``$+$'', the electron in
$K'$ is decelerating and its velocity is reversed quickly if initially $\mathbf v$ is not
exactly parallel to the direction given by Eq.~\eqref{beta}. Note that vectors $\mathbf E$,
$\mathbf{E \times B}$, $\mathbf{E \times [E \times B]}$ form an orthogonal basis thus
Eqs.~\eqref{betaEB}--\eqref{betaEEB} are enough to determine $\mathbf v$ unambiguously.

\subsection{Asymptotic trajectory}

Considering the electron motion on a timescale of the field variation timescale, $t \sim 1
\gg \tau_{\mathbf v}$, one can neglect the dynamics of the electron while it is approaching the
constant-field-approximation asymptotic solution, and assume that in every time instant the electron velocity
is determined by Eq.~\eqref{eta} or Eq.~\eqref{beta} which depend only on the instant (and local) fields.
Thus, the electron trajectory is governed by the following reduced-order equations:
\begin{eqnarray}
	\label{r}
     \frac{d\mathbf r}{dt} = \mathbf v, \\
	\label{v}
    \mathbf{E - v (v E) + v \times  B} = 0,
\end{eqnarray}
where the last equation determines the velocity field $\mathbf v$ and can be used in both B- and
E-cases (in B-case it yields Eq.~\eqref{eta}). From here on we call the solution of
Eqs.~\eqref{r}--\eqref{v} ``asymptotic trajectory'' because, first, locally it corresponds to
the asymptotic ($t \to \infty$) electron trajectory in the constant-field-approximation, and, second, it
describes the electron trajectory in asymptotically strong field ($a_0 \gg a_0^*$).

Note that the reasoning about the electron trajectory in the radiation-dominated regime is also
valid if the parameter $\chi$ is large ($\chi \approx \gamma F_\perp / e E_s$,
see Ref.~\cite{Berestetskii82, Elkina11}, where $F_\perp$ is the component of the Lorentz force perpendicular to
the particle velocity). In this case ($\chi \gg 1$) the synchrotron emission is described by
the quantum formulae and Eq.~\eqref{Frr} is not valid, however, it is still possible to describe the
electron trajectory classically between the photon emission events~\cite{Baier98, Berestetskii82} because $\ell_f \ll \ell_W$. Here
$\ell_f \sim mc^2 / F_\perp$ is the radiation formation length, i.e. the distance within which the emission of
a single photon occurs, and $\ell_W \sim c / W$ is the mean distance that the electron passes
without the photon emission; $W$ is the full probability rate of the photon emission. Estimating
\begin{equation}
    W \sim \frac{m c e^2}{\hbar^2} \frac{\chi^{2/3}}{\gamma},
\end{equation}
we obtain $\ell_f / \ell_W \sim \alpha / \chi^{1/3} < 1/137 \ll 1$.
Therefore, the electron moves
classically between the short events of the photon emission. Note also that for optical
frequencies $\ell_W / \lambda \sim \hbar \omega / (\alpha \chi^{2/3} mc^2) \ll 1$.

\section{Simple examples}
\label{tests}

In order to test the asymptotic description of the electron trajectory (Eqs.~\eqref{r} and \eqref{v}) we
compare numerical solutions of them with numerical solutions of the classical equations of the
electron motion with the radiation reaction taken into account by the inclusion of the
Landau--Lifshitz force~\cite{LandauII} or
by the recoil of the emitted photons described in the quasiclassical framework of
Baier--Katkov~\cite{Berestetskii82, Baier98}.
Numerical solution of the full equations of the electron motion is based on the Vay's
pusher~\cite{Vay08} where the Landau--Lifshitz force is taken into account with the Euler's
method or, alternatively,
the quantum recoil is taken into account by 
the Monte Carlo (MC) technique similarly to the QUILL~\cite{QUILL, Nerush17} code (see also
Appendix~\ref{appendix-tests}). In order to
solve Eqs.~\eqref{r}--\eqref{v} we use the classical Runge--Kutta method. The test results for
various field configurations are present below.

\subsection{Rotating electric field}

In the rotating electric field of the amplitude $a_0$ Eq.~\eqref{v} gives $\mathbf{ v = -E} / E$,
that coincides with the high-field limit ($a_0 \gg a_0^*$) of the Zel'dovich's stationary
solution~\cite{Zeldovich75} utilizing the main term of the Landau--Lifshitz force. This stationary
solution can be updated by taking into account quantum corrections to the radiation-reaction
force~\cite{Bashinov17}, that also yields $\mathbf{v \to \mathbf -E} / E$ in the high-field limit.
MC simulations demonstrate the same behavior, however, high dispersion of the angle between $\mathbf
v$ and $\mathbf E$ is evident, see Fig.~\ref{circe}.

\subsection{Static B-node}

\begin{figure}
	\includegraphics[width=1\linewidth]{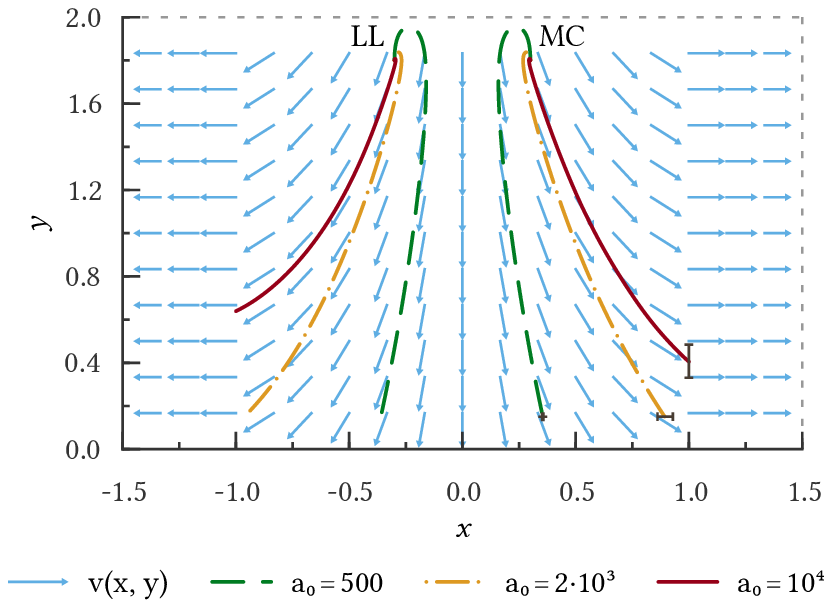}
    \caption{\label{betafield}Velocity field Eq.~\eqref{v} (arrows) and the full electron trajectories in
    the fields Eq.~\eqref{const_E_lin_B} for different values of the field magnitude: $a_0 = 500$
    (dashed lines), $a_0 = 2 \times 10^3$ (dash-dotted lines) and $a_0 = 1 \times 10^4$ (solid
    lines). The electrons start from $x_0 = \pm 0.3$ with its Lorentz factor $\gamma_0 = 100$ and
    the momentum along $y$ axis.  The trajectories at $x < 0$ are computed with the Landau--Lifshitz
    radiation reaction taken into account, while the trajectories at $x > 0$ are computed with
    radiation reaction taken into account by Monte Carlo technique and quantum formulae
    Eq.~\eqref{w}. Bars depict the standard deviation ($\pm \sigma$) of the final electron position
    computed with 400 trajectories.}
\end{figure}

Let us start from the following simple field configuration:
\begin{equation}
    \label{const_E_lin_B}
    E_y = a_0, \quad B_z = a_0 x,
\end{equation}
and the other components of the fields are zero.

In Fig.~\ref{betafield} the velocity field Eq.~\eqref{eta_explicit} ($|x| > 1$) and Eqs.~\eqref{betaEB} and
\eqref{betaE} ($|x| \leq 1$) is depicted by the arrows.
In the left half of Fig.~\ref{betafield} the electron trajectories computed with the
Landau--Lifshitz
force are shown, and in the right half of Fig.~\ref{betafield} the electron trajectories are
computed with Monte Carlo technique and quantum synchrotron formulae. Obviously, the shape of the electron trajectories computed
with Monte Carlo approach is slightly different for different runs, so the bars depict the standard deviation of
the electron final position. The trajectories are computed for different $a_0$ values, namely $a_0
= 500$, $2 \times 10^3$, $1 \times 10^4$ which correspond to dashed, dash-dotted and solid lines,
respectively. First, it is seen that at higher $a_0$ values the real electron velocity coincides
better with the velocity field that induces the asymptotic trajectories. Second, the
Landau--Lifshitz approach
demonstrate slightly better coincidence, because in the Landau--Lifshitz approach the mean electron energy generally
less than in the quantum approach.

Note that the fields Eq.~\eqref{const_E_lin_B} resemble the $B$ node of a standing linearly polarized
wave, however, in the linearly polarized standing wave the sign of $\mathbf{E \times B} |_x$ varies
in time, and the node
attracts the asymptotic electron trajectories during a half of a period, and repels them during the other
half.

\subsection{Linearly polarized standing wave}

In the linearly polarized standing wave asymptotic electron trajectories can be found analytically.
The fields of the linearly polarized standing wave read as follows:
\begin{eqnarray}
    \label{lpE}
    \mathbf E = \mathbf{y_0} a_0 \cos (t) \cos (x), \\
    \label{lpB}
    \mathbf B = \mathbf{z_0} a_0 \sin (t) \sin (x),
\end{eqnarray}
where $\mathbf y_0$ and $\mathbf z_0$ are the unit vectors along the $y$ and $z$ axes,
respectively.
Then from Eqs.~\eqref{eta_explicit}, \eqref{betaEB} and \eqref{betaE} we get:
\begin{equation}
\mathbf v = \begin{cases}
\mathbf{x}_0 \tg (t) \tg (x) \pm \mathbf{y}_0 \sqrt{1-\tg^2(t) \tg^2(x)}, & E>B \\
\mathbf{x_0} \ctg (t) \ctg (x), & E<B.
\end{cases}
\end{equation}

Since the fields are homogeneous along the $y$ axis electron's motion along it is not of any interest.
Then $x(t)$ of the asymptotic trajectory is found from the following algebraic equations:
\begin{equation}
\label{xt_lpw}
\begin{cases}
\sin (x) \cos(t) = \sin (x_0) \cos (t_0), & E>B \\
\cos (x) \sin(t) = \cos (x_0) \sin (t_0), & E<B.
\end{cases}
\end{equation}
where the starting point $x_0 = x(t_0)$ also belongs to the region $E > B$ or $E < B$. For
instance, the electron trajectory initially is determined by the first of Eqs. (\ref{xt_lpw}), then it reaches the point
with $E=B$; after that the trajectory is determined by the second of Eqs. (\ref{xt_lpw}) up to the
moment when the electron reaches another point with $E=B$ and so on.
For $E = B$ Eqs.~\eqref{lpE} and \eqref{lpB} yields
\begin{equation}
    \label{E=B}
    |\tg(x)\tg(t)| = 1
\end{equation}
with the following solution:
\begin{equation}
    x = \pm t + \frac{\pi}{2} + \pi n, \quad n = 0, \pm 1, \pm 2, ...
\end{equation}

\begin{figure}
    \includegraphics[width=1\linewidth]{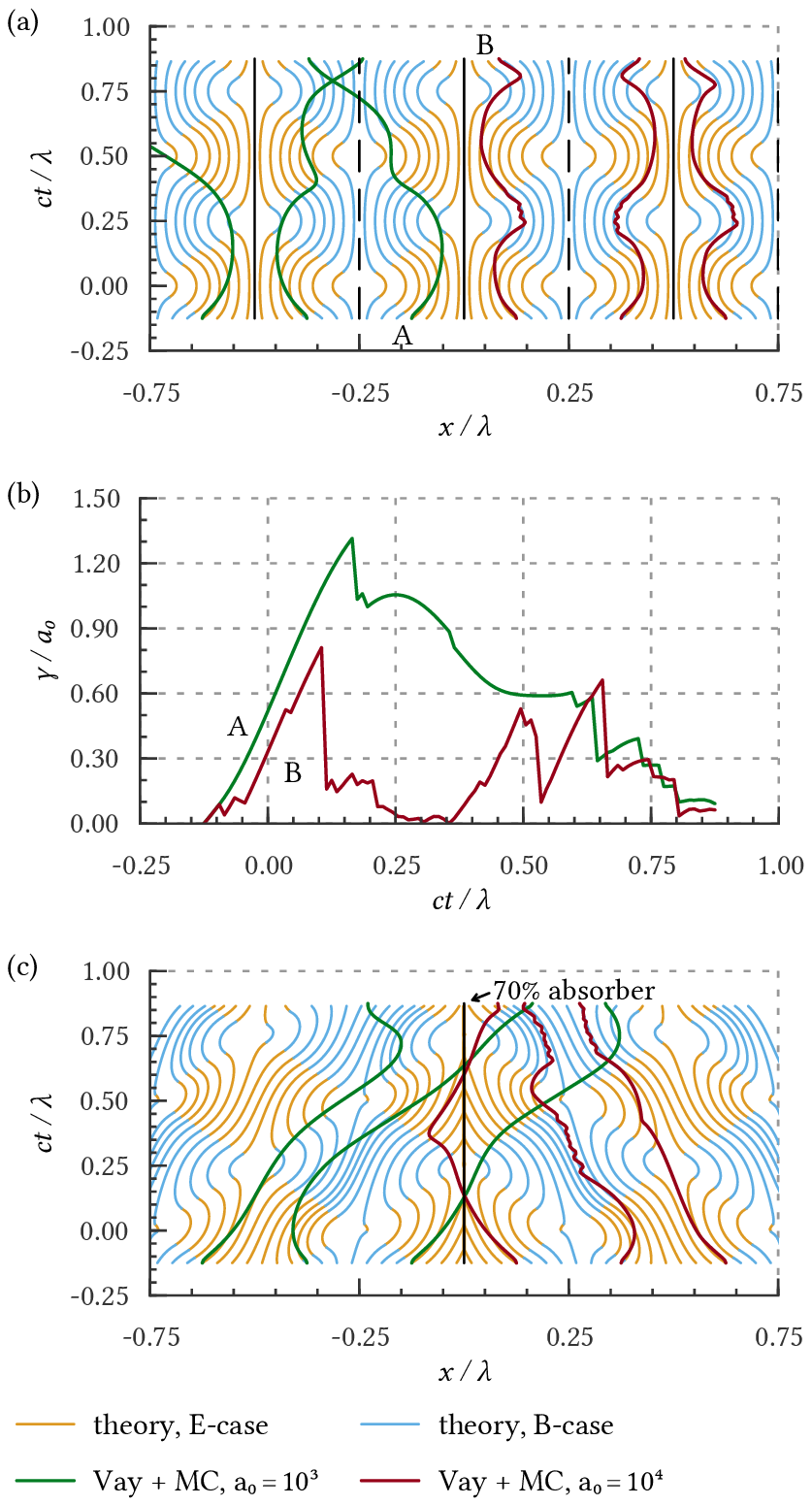}
    \caption{\label{onion} The electron motion (a), (b) in the field of the linearly polarized standing
    electromagnetic wave Eqs.~\eqref{lpE}--\eqref{lpB} and (c) in the field of two counter-propagating linearly
    polarized waves with a plane at $x = 0$ absorbing 70\% of the incoming energy (see
    Eqs.~\eqref{lpaE}--\eqref{lpaB}, $R = 0.55$).  Thin lines depict asymptotic trajectories
    obtained by
    the numerical integration of Eqs.~\eqref{r} for the E- and B-cases (beige and pale blue, respectively).
    Thick lines correspond to the numerical
    integration of the classical electron motion equations with quantum radiation
    reaction~\eqref{w} taken into account by Monte Carlo technique. It is worth to mention that the thin
    lines in (a) coincides with the analytical solution Eq.~\eqref{xt_lpw} and with the thin green lines
    in Fig.~2 from Ref.~\cite{Gonoskov14}.}
\end{figure}

For the electron starting from the point $x_0$ at the moment $t_0=0$ the chain of points $(x_1,
t_1), \; (x_2, t_2), ...$ at which $E = B$ is the following. First, from Eqs.~\eqref{xt_lpw} and
\eqref{E=B} under
the assumption that initially $E > B$, we have:
\begin{equation}
\ctg x_1 = \tg t_1 = \sqrt{\frac{1}{\cos (x_0)} - 1}
\end{equation}
The coordinate $x_2$ can be found from the observation that $x_2 = x_1$ and $t_2 = \pi - t_1$ obey
the second of Eqs.~\eqref{xt_lpw} with $x_0$, $t_0$ replaced by $x_1$, $t_1$. Also, $x_2 = x_1$ and
$t_2 = \pi - t_1$ obey Eq.~\eqref{E=B} (because $x_1$ and $t_1$ obey them).
Analogously, $x_3 = x_2$ and $t_3 = \pi + t_1$. Then the electron trajectory periodically repeat
itself (see Fig.~\ref{onion} (a)).

Note that the electron trajectory in the linearly polarized standing wave is periodic in
the framework of the presented asymptotic theory. As shown in Sec.~\ref{standing-waves}, this is just an example
of the general behaviour of the electron trajectories in standing waves in the radiation-dominated
regime. However, it can seem that this
behavior contradicts the anomalous radiative trapping~\cite{Gonoskov14} (ART). Really, ART
is caused by a drift of the electron between the asymptotic
trajectories given by Eqs.~\eqref{xt_lpw}. This drift takes many periods of the
field~\cite{Gonoskov14} and can not be described by the presented asymptotic theory.

The asymptotic electron trajectories computed with Eqs.~\eqref{r} and \eqref{v} are shown in
Fig.~\ref{onion} (a) with pale blue and beige lines (the computed trajectories coincide exactly
with the analytical solutions Eqs.~\eqref{xt_lpw}). Six electron trajectories computed with Vay's
pusher and Monte Carlo technique for the photon emission are also depicted: for $a_0 = 1 \times 10^3$ by
green lines, and for $a_0 = 1 \times 10^4$ by red lines. Fig.~\ref{onion} (b) shows the energy of the
electrons on the trajectories A and B from Fig.~\ref{onion} (a). The coincidence of the electron
trajectories computed for $a_0 = 1 \times 10^4$ with the asymptotic trajectories are evident,
opposite to $a_0 = 1 \times 10^3$ case in that the condition $\gamma \ll a_0$ is not fulfilled.
Note that in the case of $a_0 = 1 \times 10^4$ the electrons moving according to the MC approach to
the radiation reaction, become closer to the
B-nodes for each subsequent period, that is the effect of ART.

\section{Absorption-induced trapping}
\label{absorption-induced-trapping}

It follows from Eqs.~\eqref{eta_explicit} and \eqref{betaEB} that the angle between the asymptotic velocity
$\mathbf v$ and the Poynting vector $\mathbf{ S \propto E \times B}$ is always less than $\pi /
2$, i.e. $\mathbf{ v \cdot S} > 0$. This hints that the electron motion in the radiation-dominated
regime can be connected with the energy flow of the electromagnetic fields. Let us
consider the region containing currents which (partially) absorb the incoming electromagnetic wave.
In the average, the Poynting vector is directed into the region of the currents, and we suggest that
in the radiation-dominated regime this region attracts the electron trajectories.
In this section we verify this suggestion in a couple of examples.

A plane wave pushes initially immobile electrons approximately in the direction of the Poynting
vector, so it can seem that the absorption-induced trapping can be realized without strong radiation
reaction. However, as seen from the examples below, in the absence of strong radiation losses if the
electrons have been accelerated by a wave, then they can not be turned back by a counter-propagating
wave. Thus the radiation reaction may cause electron trapping in the region with strong absorption
of the electromagnetic energy.

\subsection{Counter-propagating linearly polarized waves partially absorbing by a plane}

The field of two counter-propagating linearly polarized (along the $y$ axis) waves, that is
partially absorbing at the
plane $x = 0$, can be written as follows:
\begin{eqnarray}
    \label{lpaE}
    \mathbf E = \mathbf{y_0} a_0 \left\{ \cos (t) \cos (x) - 0.5 (1 - R) \cos(x \mp t) \right\}, \\
    \label{lpaB}
    \mathbf B = \mathbf{z_0} a_0 \left\{ \sin (t) \sin (x) \mp 0.5 (1 - R) \cos(x \mp t) \right\},
\end{eqnarray}
where in $\mp$ the upper sign corresponds to $x > 0$ and the lower one corresponds to $x < 0$, and
$R$ is the reflection coefficient. The asymptotic and Vay+MC electron trajectories in this field are shown in
Fig.~\ref{onion} (c) with the same color codes as in Fig.~\ref{onion} (a). Here $R = 0.55$ that
means absorption of $70 \%$ of the wave energy in the plane $x = 0$. As seen from the figure, the
electron trajectories are attracted by the plane $x = 0$ in the strong radiation-dominated regime, whereas
at moderate intensity of the waves the electrons easily pass the plane. The mean standard
deviation of $x$ computed for the Vay+MC electron trajectories for ten periods of the wave and $x_0 =
0.25 \lambda$ is about
$1.5 \lambda$ for $a0 = 1 \times 10^3$ and $0.2 \lambda$ for $a_0 = 1 \times 10^4$.

\subsection{Multipole wave absorbing by a current loop}

\begin{figure}
	\includegraphics[width=1\linewidth]{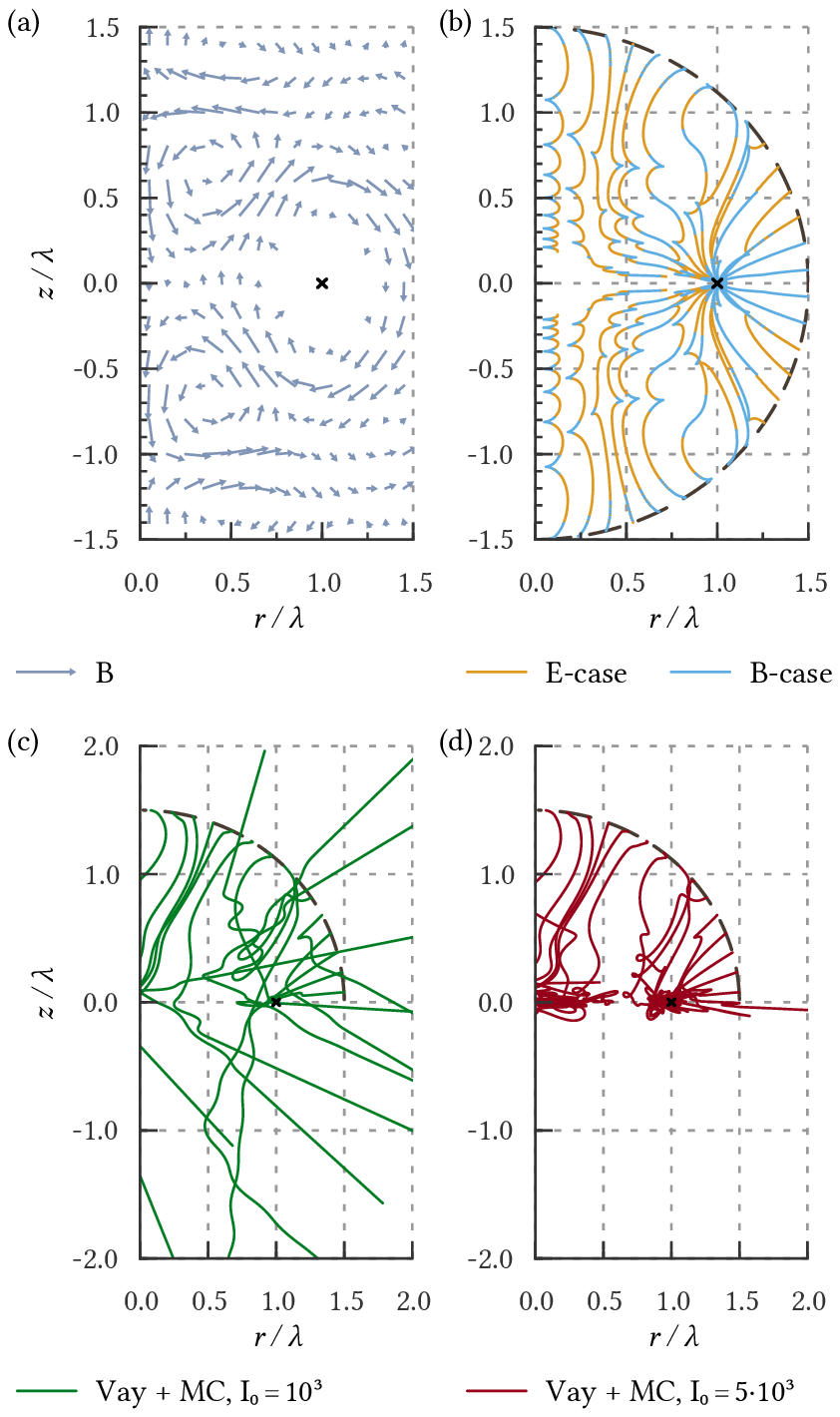}
    \caption{\label{currentloop} (a) The magnetic field of a multipole wave that is entirely
    absorbing by a current loop (see App.~\ref{appendix-multipole-wave}), the loop radius is $r_\ell = \lambda = 1 \text{ } \mu
    \text{m}$, $t = 0$. The axis of the loop coincides with the $z$ axis and the position of a ``wire''
    is shown by the black cross. (b) Asymptotic electron trajectories for E- and B-case (orange and
    blue, respectively) in the multipole wave. The electrons start to move at $t = 0$ from the
    points on the circle $(r^2 + z^2)^{1/2} = 1.5 \, r_\ell$ (thick black dashed line). The bottom plots
    show the Vay+MC electron trajectories in the field of the multipole wave for (c) $I_0 = 1 \times
    10^3$ and (d) $I_0 = 5 \times 10^3$. All trajectories are computed for $t \in [0, 5 \lambda /
    c]$.}
\end{figure}

The field of a multipole harmonic wave that is completely absorbing by a current loop can be obtained
by time reversal of the field emitting by a current loop (see
App.~\ref{appendix-multipole-wave}). The
electron motion in the absorbing multipole wave with the angular frequency $\omega = 2 \pi c / \lambda$
for a loop radius $r_\ell = \lambda = 1 \; \mu \text{m}$ is shown in Fig.~\ref{currentloop},
where in the cylindrical coordinate system the ``wire'' position is marked with the black cross.
The $z$ axis is the axis of the loop.
Fig.~\ref{currentloop} (a) demonstrates the magnetic field of the multipole wave at $t = 0$. Fig.~\ref{currentloop} (b)
shows the asymptotic electron trajectories, Figs.~\ref{currentloop} (c) and (d) show the electron
trajectories computed by Vay and MC
algorithm for the loop current magnitude $I_0 = 1 \times 10^3$ and for
$I_0 = 5 \times 10^3$, respectively. The trajectories start at $t = 0$ from the sphere shown by a thick dashed
line and are computed up to $t = 5 \lambda / c$. 

It is seen from Fig.~\ref{currentloop} that the current loop attracts
the asymptotic electron trajectories. However, it is seen that the absorption-induced trapping is
not really a strict trapping but
just means that electrons in the radiation-dominated regime stay for a long time in the region with
the currents absorbing
the incoming waves.

\section{Emission-absorption symmetry and general properties of asymptotic trajectories}
\label{properties}

In this section we consider the properties of the electron trajectories described by Eqs.~\eqref{r}
and \eqref{v}. For this purpose let us consider the well-known symmetry of the Maxwell's equations,
namely, the following transform
\begin{eqnarray}
    \label{symmetry_t}
    t^* = - t, \\
    \label{symmetry_E}
    \mathbf{E^* = -E}, \\
    \label{symmetry_B}
    \mathbf{B^* = B}, \\
    \label{symmetry_j}
    \rho^* = -\rho, \qquad \mathbf{j^* = j}.
\end{eqnarray}
does not change them, i.e. they leads to the Maxwell's equations for the starred variables; here
$\rho$ is the charge density and $\mathbf j$ is the current density. From
here on we denote $\mathbf{E}$, $\mathbf{B}$, $\mathbf{j}$ evolving in time $t$ as {\it initial}
system and $\mathbf{E}^*$, $\mathbf{B}^*$, $\mathbf{j}^*$ evolving in time $t^*$ as {\it starred}
system. This symmetry is the relation between a system of currents emitting some fields and the
system of currents absorbing the fields: namely, the Poynting vector, the $\mathbf{j \cdot E}$
product and the time direction in the starred system is opposite to that in the initial system.

According to~Eq.~\eqref{v}, in the starred system the velocity field $\mathbf{v}^*$ relates to the velocity
field of the initial system $\mathbf{v}$ as follows:
\begin{equation}
    \label{symmetry_v}
    \mathbf{v}^*(\mathbf{r}, t^*) = -\mathbf{v(\mathbf{r}, -t^*)},
\end{equation}
that obeys the stability condition $\mathbf{v^* \cdot E^*} < 0$. Thus, in the starred system
the velocity field and the time direction are opposite to that in the initial system, that
leads to the same trajectories in the starred system $\mathbf{r}^*(t^*)$ as in the initial system,
passed by the electrons in the opposite direction:
$d \mathbf{r}^* / dt^* = \mathbf{v}^*(\mathbf{r}^*, t^*) = -\mathbf{v}(\mathbf{r}^*, -t^*)$.

Let us note a fundamental difference between the asymptotic trajectories described by Eqs.~\eqref{v}, \eqref{r},
and by the ponderomotive description.
The ponderomotive force is determined by the distribution of $E^2$ and $B^2$ and is
indifferent to the transform Eqs.~\eqref{symmetry_t}--\eqref{symmetry_j}, whereas this transform reverses
the direction of the electron motion in the case when Eqs.~\eqref{v} and \eqref{r} are applicable,
namely, when radiation reaction is strong.

In order to illustrate the difference between the ponderomotive description and the description by the velocity
field Eq.~\eqref{v} the following toy example can be considered. The first laser pulse propagates along
the direction $\mathbf{x}_0$ and scatters an electron aside. Then
the second pulse is formed from the first one with the
substitution~\eqref{symmetry_E}--\eqref{symmetry_B}, and according to the Maxwell's equations this pulse
travels in the direction $-\mathbf{x}_0$. In the framework of the ponderomotive description the
second laser pulse is not important
because it will never meet the electron scattered by the first pulse. At the same time, considering
the first laser pulse as
the initial system of fields, we see that the second laser pulse is
equivalent to the starred system of fields. In the case of strong
radiation reaction the asymptotic approach is valid, and the electron according to
Eq.~\eqref{symmetry_v} will pass along its preceding trajectory in the opposite direction in the
field of the second pulse, i.e.
the electron will be brought back to its initial position by the second laser pulse.

Therefore, the asymptotic description of the electron motion~Eqs.~\eqref{r} and \eqref{v} implies that
the electrons are not scattered by, but stay for a long time in the field of a laser pulse or in a
laser beam. This conclusion is in a good agreement with the results of theoretical considerations
and numerical simulations showing that the ponderomotive force can be significantly suppressed by the
radiation reaction~\cite{Fedotov14b, Ji14b}.

\subsection{Asymptotic trajectories in standing waves}
\label{standing-waves}

We see in Sec.~\ref{tests} that the reduced equations lead to periodic electron trajectories in
the linearly polarized standing electromagnetic wave.
Here we show, that Eqs.~\eqref{r} and \eqref{v} always lead to a periodic electron trajectories in a
wide class of fields, namely in the \textit{periodic} fields which can be represented in the following form:
\begin{eqnarray}
    \label{Esw}
    \mathbf{E} = \mathbf{f}(\mathbf{r}, t) - \mathbf{f}(\mathbf{r}, -t), \\
    \label{Bsw}
    \mathbf{B} = \mathbf{g}(\mathbf{r}, t) + \mathbf{g}(\mathbf{r}, -t),
\end{eqnarray}
where $\mathbf{E = f}(\mathbf{r}, t)$, $\mathbf{B = g}(\mathbf{r}, t)$ is the solution of Maxwell's
equations for some charge density $\rho$ and current density $\mathbf{j}$ (for the sake of simplicity let
us consider $\rho = 0$ and $\mathbf{j} = 0$). This representation
means that the fields are the sum of the fields of some system and the fields of the corresponding
starred system. In this case the symmetry \eqref{symmetry_t}--\eqref{symmetry_j} leads to the same
fields of the starred system as in the initial system, i.e. $\mathbf{E}^*(\mathbf{r}, t^*) = \mathbf{E}(\mathbf{r}, t^*)$, $\mathbf{B}^*(\mathbf{r}, t^*)
= \mathbf{B}(\mathbf{r}, t^*)$, hence it should lead to the same
velocity field $\mathbf{v}^*(\mathbf{r}, t^*) = \mathbf{v}(\mathbf{r}, t^*)$, that together with
Eq.~\eqref{symmetry_v} yields
\begin{equation}
    \mathbf{v}(\mathbf{r}, -t) = -\mathbf{v(\mathbf{r}, t)}.
\end{equation}
Thus, the velocity field in the electromagnetic fields~\eqref{Esw}--\eqref{Bsw} is an odd function of
time. Consequently, the time reversal conserves the equation for the electron position,
\begin{equation}
    \frac{d \mathbf{r}}{d(-t)} = \mathbf{v}(\mathbf{r}, (-t)),
\end{equation}
and the electron position $\mathbf{r}(t)$ is an even function of time. Therefore,
\begin{multline}
    \mathbf{r}(t) - \mathbf{r}(-t) = \int_{-t}^t \mathbf{v}(\mathbf{r}(t), t) \, dt = \\
    \int_{0}^t \mathbf{v}(\mathbf{r}(t), t) \, dt + \int_{0}^t \mathbf{v}(\mathbf{r}(-t), -t) \, dt
    = 0.
\end{multline}

As far as the velocity field governed by Eqs.~\eqref{eta_explicit} and \eqref{betaB}--\eqref{betaEEB} is
a single-valued function of the electromagnetic fields, and the fields are periodic in time, the
velocity field is also periodic with the same period, $T$. Thus, the velocity field is an odd
function relative to any
time instant $t = n T$, where $n$ is an integer. Let the electron starts to move at $t = nT - T /
2$, then it comes to the starting point a period later, $\mathbf{r}(nT + T / 2) = \mathbf{r}(nT - T /
2)$, then due to the periodicity of $\mathbf{v}$ at $t = (n + 1) T$, we have $\mathbf{r}((n + 1)T + T /
2) = \mathbf{r}(nT + T / 2)$. Therefore, in the framework of the asymptotic approach, the electron is moving periodically back and forth along
the same path in the periodic fields Eqs.~\eqref{Esw}--\eqref{Bsw}.

\subsection{Asymptotic trajectories in a laser beam of finite diameter}
\label{TE11}

\begin{figure*}
    \centering
	\includegraphics{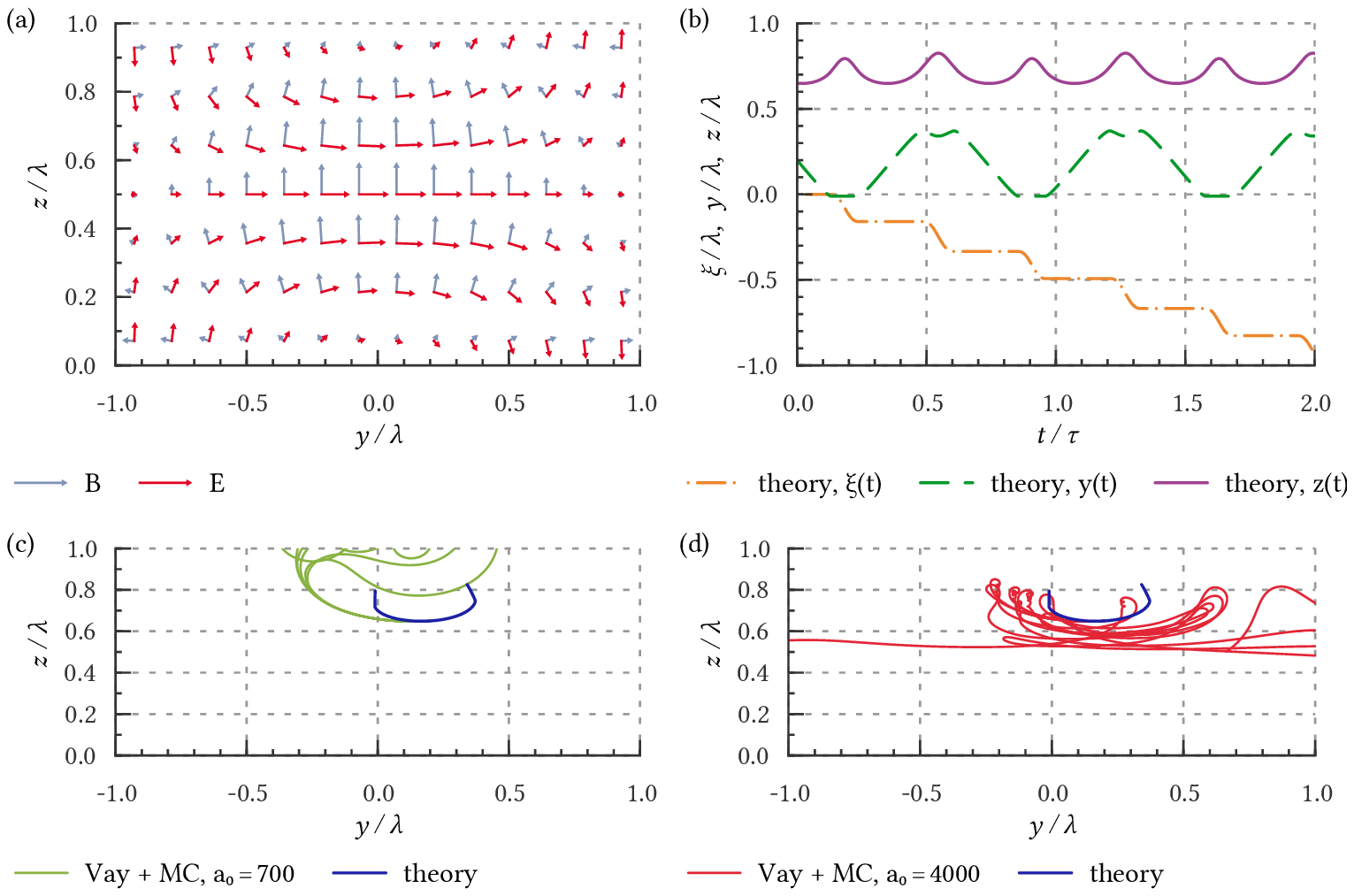}
    \caption{\label{te11}(a) The electric and magnetic fields (red and blue arrows, respectively) of the continued TE11 mode of a
    waveguide, Eqs.~\eqref{te11-2}--\eqref{te11-3} and \eqref{te11-5}--\eqref{te11-6}, at $t = 0$.
    (b) The asymptotic
    trajectory of an electron computed with Eqs.~\eqref{r}, \eqref{eta_explicit} and
    \eqref{betaB}--\eqref{betaEEB} in the laboratory reference frame; $\xi = x - v_g t$, where
    $v_g$ is the group velocity of the TE11 mode.
    (c), (d) The same asymptotic trajectory (thick line) and five full electron trajectories (thin
    lines) starting at the point $x = 0$, $y = 0.2 \, \lambda$, $z = 0.65 \, \lambda$ ($\lambda = 1
    \text{ } \mu \text{m}$, $t \in [0, 2 \tau]$), for (c) $a_0 = 700$ and (d) $a_0 = 4 \times
    10^3$.}
\end{figure*}

Here we stress that many field configurations could be reduced to the form
of periodic fields that obey the emission-absorption symmetry
Eqs.~\eqref{symmetry_t}--\eqref{symmetry_j}. In the previous subsection we also assumed that the
velocity field is a single-valued function of the fields. This is not strictly true in the B-case,
because one can add to $\mathbf{v}$ from Eq.~\eqref{eta_explicit} a vector parallel to $\mathbf{B}$.
The effect of this ambiguity is also discussed in this section.

Let us consider the fields of TE11 mode of a rectangular metallic waveguide:
\begin{eqnarray}
    \label{te11-1}
    E_x = 0, \\
    \label{te11-2}
    E_y = a_0 \cos(k_y y) \sin(k_z z) \cos(t - k_x x), \\
    \label{te11-3}
    E_z = -\frac{a_0 k_y}{k_z} \sin(k_y y) \cos(k_z z) \cos(t - k_x x), \\
    \label{te11-4}
    B_x = \frac{a_0 (k_z^2 + k_y^2)}{k_z} \cos(k_y y) \cos(k_z z) \sin(t - k_x x), \\
    \label{te11-5}
    B_y = -k_x E_z, \\
    \label{te11-6}
    B_z = k_x E_y,
\end{eqnarray}
where we assume that the wave angular frequency $\Omega = (k_x^2 + k_y^2 + k_z^2)^{1/2}
= 1$ (here we use the normalization frequency $\omega$ equal to the frequency of the wave, and, as before, the
time is normalized to $1/\omega$, coordinates are normalized to $c / \omega$, $\mathbf k$ is the
wavenumber normalized to $\omega / c$).
These fields obey the metallic boundary conditions at $y = 0, \; \pm \ell_y, \; \pm 2 \ell_y,...$
($E_z = 0$) and at $z = 0, \; \pm \ell_z, \; \pm 2 \ell_z,...$ ($E_y = 0$). Here $\ell_y = \pi /
k_y$ and $\ell_z = \pi / k_z$ are the sizes of the waveguide along the $y$- and $z$-axes,
respectively.

The fields Eqs.~\eqref{te11-1}--\eqref{te11-6} are the solution of the Maxwell's equations not only inside the waveguide but in
the open space as well because this fields can be represented as a sum of plane waves. Particularly, we consider these fields in the region $y \in [-\ell_y / 2, \ell_y /
2]$ and $z \in [0, \ell_z]$ as the model of the laser beam of finite diameter. If $\ell_y \gg
\ell_z$, the electric field is mainly directed along the $y$-axis and reaches its maximum in the
center of the beam.

The fields Eqs.~\eqref{te11-1}--\eqref{te11-6} are shown
in Fig.~\ref{te11} (a) for $\ell_y = 4 \pi$, $\ell_z = 2 \pi$ and $t = x = 0$. The asymptotic electron
trajectory computed for these fields is shown in Fig.~\ref{te11} (b), where $\xi = x - v_g t$, $v_g
= k_x \approx 0.83$ is the group velocity of the electromagnetic wave, and the trajectory starts at
$t = 0$, $x = 0$, $y = 0.2$ and $z = 0.65$ and is computed up to $t = 2 \tau$, where
\begin{equation}
    \label{tau}
    \tau = \frac{2 \pi}{k_x (v_\phi - v_g)} = \frac{2 \pi}{1 - k_x^2}
\end{equation}
is the intrinsic timescale of the task, $v_\phi = 1 / v_g$ is the phase
velocity of the wave. For Fig.~\ref{te11} $c \tau / \lambda \approx 3.2$.
In Figs.~\ref{te11} (c) and (d) the same asymptotic trajectory is shown as the thick blue line. Five
electron trajectories in the fields given by Eqs.~\eqref{te11-1}-\eqref{te11-6} are shown in Figs.~\ref{te11}
(c) and (d). These trajectories start at the same point as the asymptotic trajectory, but they are
computed by Vay's algorithm with quantum radiation reaction incorporated by Monte Carlo technique.
For Figs.~\ref{te11} (c) and (d) we use $a_0 = 700$ and $a_0 = 4 \times 10^3$, respectively.

It is seen from Figs.~\ref{te11} (b), (c) and (d) that the asymptotic trajectory is quasiperiodic,
that is in a qualitative agreement with the fact that
for $a_0 = 4 \times 10^3$ the electrons stay for a long time in the high-field region. However, as
we see below, the asymptotic trajectories being computed in
the laboratory reference frame yield the values of $\xi$ and the values of the trajectory period
which do not coincide well with that for real electron trajectories. The reason for that is that
Eq.~\eqref{eta_explicit} is not Lorentz invariant, namely if one
compute $\mathbf v$ from it in some reference frame, in another reference frame he obtain ${\mathbf
v' = E' \times B' + a B'}$, where $a$ is a constant.

Let us transform the fields Eqs.~\eqref{te11-1}--\eqref{te11-6} to the reference frame $K'$ moving
along the $x$ axis with the group velocity of the fields $v_g$. This lead to the following
fields:
\begin{eqnarray}
    \label{te11'-1}
    E_y' = a_0 k_\perp \cos(k_y y) \sin(k_z z) \cos(k_\perp t'), \\
    \label{te11'-2}
    E_z' = -\frac{a_0 k_\perp k_y}{k_z} \sin(k_y y) \cos(k_z z) \cos(k_\perp t'), \\
    \label{te11'-3}
    B_x' = \frac{a_0 (k_z^2 + k_y^2)}{k_z} \cos(k_y y) \cos(k_z z) \sin(k_\perp t'), \\
    \label{te11'-4}
    E_x' = B_x' = B_y' = B_z' = 0,
\end{eqnarray}
where $k_\perp = \sqrt{1 - k_x^2}$. These fields do not depend on $x'$ and for all the electrons in
these fields the component of the
Lorentz force along the $x'$ axis is absent. Furthermore, the electrons due to the radiation
reaction ``forget'' their initial direction of motion, hence we conclude that the average velocity
of the electrons in the fields Eqs.~\eqref{te11'-1}--\eqref{te11'-4} is $v_x' = 0$. Therefore, in $K$
the average electron velocity is $v_x = v_g$ hence $\xi = \operatorname{const}$ that is in good agreement with results of Vay+MC
simulations. Note that $\xi = \operatorname{const}$ does not coincide with
the result of the asymptotic consideration in the laboratory reference frame (see Fig.~\ref{te11}
(b)). Also, a wrong value of $v_x$ leads to a wrong value of the period of $y$ and $z$ coordinates of
the electron in the framework of the asymptotic approach.

The substitution $t' \rightarrow t' + \pi / 2 k_\perp$ yields that the electric field
given by Eqs.~\eqref{te11'-1}--\eqref{te11'-2} are odd functions of time and the magnetic field
Eq.~\eqref{te11'-3} is the even function of time in $K'$. As follows from Sec.~\ref{standing-waves}
in this case the electron trajectories are periodic in the radiation-dominated regime and their
period is equal to $2 \pi / k_\perp$ in $K'$. Therefore,
in the laboratory reference frame in the radiation-dominated regime the electrons move along the $x$-axis
with the group velocity of the laser beam, and, as $y' = y$ and $z' = z$, the electron trajectories
are periodic in the $yz$ plane with the period
\begin{equation}
    \frac{2 \pi}{k_\perp \sqrt{1 - v_g^2}} = \tau.
\end{equation}
Thus, the ambiguity of the velocity field in the asymptotic approach can be resolved by appropriate
choose of the reference frame.

Therefore, we show that the asymptotic description, Eqs.~\eqref{r} and \eqref{v}, leads to periodic
trajectories in a wide class of standing waves (e.g. formed by laser beams of finite diameter), and to
electron motion along the laser beam with its group velocity with periodic transverse motion. The
latter may explain the effect of the radiation-reaction trapping~\cite{Ji14b}.

\section{Conclusion}
\label{conclusion}

To conclude, here we show that in the radiation-dominated regime the electrons tend to move
with velocity that is determined by the fields only, see Eq.~\eqref{v}. This means that the
electron trajectory can be found from the first-order equation, Eq.~\eqref{r}. We call this
velocity \textit{asymptotic} because it can be found as the asymptotic electron velocity ($t \to
\infty$) in the constant
field approximation. The reason for reduction of the equation order
is that the electron energy in the
radiation-dominated regime is small ($\gamma \ll a_0$), the electrons are ``light'' and are easily
turned by the laser field to the asymptotic direction in a time much smaller than the characteristic variation time of the electromagnetic
fields.  The velocity field ${\mathbf v}({\mathbf r}, t)$ corresponds to the absence of the component of
the Lorentz force transverse to the electron velocity, so $\mathbf v$ is also called the radiation-free
direction~\cite{Gonoskov17}.

In a number of the electromagnetic field configurations we found the numerical solutions of
the reduced-order equations and the full equations of electron motion with the radiation reaction taken
into account by the Monte Carlo technique and the Baier--Katkov synchrotron formulae~\cite{Berestetskii82}. The comparison
between these solutions demonstrates that the reduced-order equations can be used for a qualitative
description of the electron trajectories for $a_0$ greater or of the order of thousand for optical
wavelengths. In order to stress these high values of $a_0$ we call the solutions of the reduced
equations of motion as asymptotic trajectories ($a_0 \to \infty$).

Also we demonstrate that the reduced-order equations for the electron trajectories in the
radiation-dominated regime are the useful analytical tool. First, they predict the electron
trapping in the regions where the wave field is absorbed, see
Sec.~\ref{absorption-induced-trapping}. This result can be important for the theoretical
consideration of the field absorption by the QED cascade in the counter-propagating laser
waves~\cite{Nerush11a}. Second, contrary to the concept of the
ponderomotive force, the asymptotic theory leads to periodic
electron trajectories in a wide class of standing electromagnetic fields (including the case of
counter-propagating tightly focused laser beams, see Sec.~\ref{standing-waves}). This result is in
a good agreement
with Ref.~\cite{Fedotov14b} that demonstrates the reduction of the ponderomotive force in the
radiation-dominated regime. Furthermore, using a certain configuration of the laser beam we demonstrate
that the beam in the radiation-dominated regime does not push the electrons aside, but captures
and carries them with the group velocity of the beam. This result probably explains the radiation-reaction
trapping observed in the numerical simulation of Ref.~\cite{Ji14b}.

Therefore, the concept of the ponderomotive force is not applicable in the radiation-dominated
regime and can be replaced by the description of the asymptotic electron
trajectories. This concept implies that velocities of the electrons in a given point
are the same hence the electrons (positrons) in the radiation-dominated regime can be described
in the framework of the hydrodynamical approach. The Maxwell's equations, in which the electron current is determined only by the plasma
density and by the local field values (see Eq.~\eqref{v}), together with the continuity equation for the
plasma density are formed the closed system of equations. Note that the reduced-order equations
gives the positive field work on the electrons
(${\mathbf v \cdot E} < 0$) hence
the plasma in the framework of the asymptotic theory is always an absorbing medium. In more
details this hydrodynamical approach will be considered elsewhere.

\begin{acknowledgments}
    We thank A.~V.~Bashinov and V.~A.~Kostin for fruitful discussions. We are grateful to
    E.~V.~Frenkel who brought our attention to the symmetries of the Maxwell's equations, and to
    T.~Docker for his help with \textit{haskell-chart} library.

    This research was supported by the Grants Council under the President of the Russian Federation
    (Grant No. MK-2218.2017.2). The study of the absorption-induced trapping was supported
    by the Russian Science Foundation through Grant No. 16–12-10383
\end{acknowledgments}

\appendix

\section{Tests of numerical instruments}
\label{appendix-tests}

\subsection{Radiation reaction: classical limit}

In order to test the Vay's solver for the equations of motion~\cite{Vay08} coupled with
Landau--Lifshitz radiation
reaction force (taken into account by Euler method) let us
consider electron motion in constant crossed electric and magnetic fields:
\begin{eqnarray}
    \label{crossed_fields1}
    E_y = a_0 / 2, \quad B_z = a_0, \\
    \label{crossed_fields2}
    E_x = E_z = B_x = B_y = 0.
\end{eqnarray}

In the reference frame $K'$ moving along $x$ axis with the speed $V = 0.5$ the electric field
vanishes and the only $z$ component of the magnetic field remains: $B_z' = B_z \sqrt{1 - V^2}$,
where the stroke marks quantities in $K'$.  Taking into account the Landau--Lifshitz radiation
reaction, for relativistic electron motion in $K'$ we obtain (assuming $\gamma \gg 1$):
\begin{eqnarray}
    \label{dg'dt'}
    d \gamma' / dt' = -C \gamma'^2, \\
    dw' / dt' = i B_z' w' / \gamma', \\
    \label{dv_z'dt'}
    dv_z' / dt' = 0,
\end{eqnarray}
where
\begin{equation}
    C = \frac{2}{3} \frac{e^2}{\hbar c} \frac{\hbar \omega}{m c^2} B_z'^2 v_\perp^2,
\end{equation}
$w' = v_x' + i v_y'$, $v_\perp^2 = v_x'^2 + v_y'^2$ and $\omega$ is just some frequency used for
normalization of time. The solution of Eqs.~\eqref{dg'dt'}-\eqref{dv_z'dt'} is the following:
\begin{eqnarray}
    \label{cornu_g}
    \gamma' = \frac{\gamma_0'}{1 + \gamma_0' C t'}, \\
    w' = w_0' \exp \left( \frac{i B_z'}{\gamma_0'} (t' + \frac{\gamma_0 C t'^2}{2})\right),
\end{eqnarray}
and
\begin{multline}
    \label{cornu_xy}
    x' + i y' = x_0' + i y_0' +
    w_0' \sqrt{\frac{i \pi}{2 B_z' C}} \exp \left(-\frac{i B_z'}{2 \gamma_0'^2 C}\right) \\
    \times \left\{ \operatorname{erf}\left(\sqrt{\frac{B_z' C}{2 i}} (t' + \frac{1}{\gamma_0'
    C})\right)
    \right. \\
    \left. - \operatorname{erf}\left(\sqrt{\frac{B_z' C}{2 i}} \frac{1}{\gamma_0' C}\right) \right\},
\end{multline}
where subscript $0$ denotes $t' = 0$ and
\begin{equation}
    \operatorname{erf}(x) = \frac{2}{\sqrt{\pi}} \int_0^x \exp(-t^2) \, dt
\end{equation}
is the error function.

\begin{figure}
	\includegraphics[width=1\linewidth]{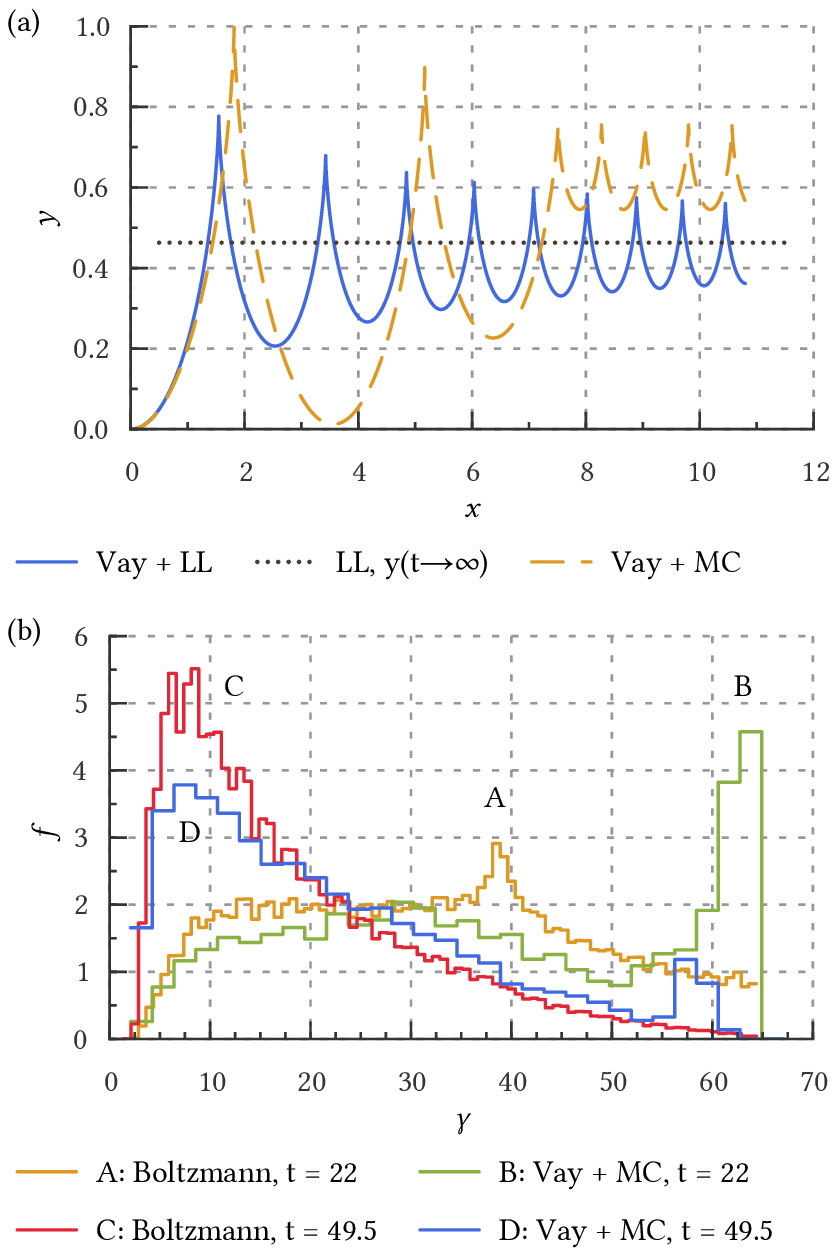}
    \caption{\label{cornu}(a) The electron trajectory in the crossed electric and magnetic
    fields~\eqref{crossed_fields1}-\eqref{crossed_fields2} ($a_0 = 50.3$, $\lambda = 1.1 \text{
        nm}$) computed by numerical integration of the classical equations of the electron motion
    with radiation reaction taken into account by means of the main term of the Landau--Lifshitz
    radiation reaction force (solid line) and
    by means of Monte Carlo technique and quantum emission probability~\eqref{w} (dashed
    line). The dotted line depicts $y(t \rightarrow \infty)$ found from Eq.~\eqref{cornu_xy}. The
    electrons initially have $v_x \simeq 0.8$, $v_z \simeq 0.6$ and $\gamma = 63$. (b) In the same
    fields, the energy distribution of the electrons with the same initial momentum, computed by
    the same method as for the dashed line in the subplot (a) (lines B and D) and with numerical
    integration of the Boltzmann equation~\eqref{Boltzmann}, for different time instants.}
\end{figure}

Figure~\ref{cornu} (a) demonstrates the electron trajectory in the $xy$ plane obtained with
the numerical integration of the equation of motion taking into account radiation reaction in
Landau--Lifshitz form (solid blue line) for $a_0 = 50.3$, $\omega = 2\pi c / \lambda$, $\lambda =
1.1 \text{ nm}$, $x_0 = y_0 = 0$, $v_x(t = 0) \simeq 0.8$, $v_z(t = 0) \simeq 0.6$ and $\gamma_0 =
63$. Note that these parameters ensure $v_\perp'(t) = v_x'(t = 0) = 0.5 = V$, leading in the
laboratory reference frame to the cycloid-like trajectory with points of $dy/dx \rightarrow
\infty$.

For the given parameters we obtain $B'_z = 43.6$, $\gamma_0' = \gamma'(t' = 0) = 43.6$ and $C = 4.9
\times 10^{-3}$, and from Eq.~\eqref{cornu_g} at the time instant $t_1' = 6 \pi$ we get
$\gamma'(t_1') = \gamma_0' / 5$. Neglecting displacement of the particle, $x'$, and assuming $v_x'(t_1')
= V$, we finally get for the laboratory reference frame: $t_1 = (1 - V^2)^{-1/2} t_1' \approx 22$ and
$\gamma(t_1) \approx 12.6$. It should be mentioned that at the time instance at which $v_x' = V$,
in the laboratory reference frame the Lorentz factor reaches its local maximum. Numerical solver using
Landau--Lifshitz force demonstrates that the local maximum of $\gamma$ closest to $t_1 = 22$ is
reached at $t \approx 21.5$ and is $\gamma \approx 13.4$ that is quite close to the predicted value.

Equation~\eqref{cornu_xy} yields $y'(t' \rightarrow \infty) \approx 0.463$ for the above-mentioned
parameters. This value ($y(t \rightarrow \infty) = y'(t' \rightarrow \infty)$) is depicted as gray
dotted line in Fig.~\ref{cornu}, and in a good agreement with the value obtained with the numerical
solver. The dashed orange line is got by means of particle pusher that takes into account with
the quantum formulae and is described in the next subsection.

\subsection{Radiation reaction: general case}

The quantum radiation reaction can be taken into account in Vay's pusher by means of Monte Carlo
technique. To do this we use the alternative event generator~\cite{Elkina11} based on Baier--Katkov
synchrotron formula~\cite{LandauII, Baier98}. The event generator checks at every time step if the photon emission
occurs, and if it does, the electron momentum is decreased on the momentum of the emitted photon. Using of classical
description of the electron trajectory together with the quantum formula for the photon emission is
valid because the radiation formation length in strong fields ($a_0 \gg 1$) is much smaller than
the field characteristic scale~\cite{LandauII, Baier98, Gonoskov15}.

In order to test Vay's pusher coupled with Monte Carlo event generator we compute the energy
distribution of the electrons in the crossed fields
Eqs.~\eqref{crossed_fields1}--\eqref{crossed_fields2}. The resulting spectra are compared with
the spectra obtained from the Boltzmann equation in the reference frame $K'$.

As mentioned above, in the reference frame $K'$ moving along $x$
axis with velocity $V = 0.5$ the electrons see the pure magnetic field directed along the $z$ axis.
Therefore, in $K'$ the Boltzmann equation that describes the electron energy distribution $f'(t',
\gamma')$ is the
following:
\begin{multline}
    \label{Boltzmann}
    \frac{\partial f'(t', \gamma')}{\partial t'} = \int_{\gamma'}^\infty w(\epsilon,
    \epsilon - \gamma') f'(t', \epsilon) \, d\epsilon \\
    - W(\gamma') f'(t', \gamma'),
\end{multline}
where
\begin{multline}
    \label{w}
    w(\epsilon, \epsilon_\gamma) = - \frac{\alpha}{\epsilon_{\ell} \epsilon^2}
      \left[\int_\varkappa^\infty \operatorname{Ai}(\xi) \, d\xi \right. \\
        \left. + \left( \frac{2}{\varkappa} + \frac{\epsilon_\gamma \chi
        \varkappa^{1/2}}{\epsilon}\right)
      \operatorname{Ai}'(\varkappa) \right],
\end{multline}
is the distribution of the photon emission probability by the electron with the Lorentz
factor $\epsilon$
over the photon energy $\varepsilon_\gamma$ normalized to $mc^2$, i.e. over $\epsilon_\gamma =
\varepsilon_\gamma / mc^2$ (see
Refs.~\cite{LandauII, Baier98}), and
\begin{eqnarray}
    \chi = \epsilon_\ell B_z' v_\perp \epsilon, \\
    \varkappa = \left[\frac{\epsilon_\gamma}{(\epsilon - \epsilon_\gamma) \chi} \right]^{2 / 3}, \\
    W(\gamma') = \int_0^{\gamma'} w(\gamma', \epsilon_\gamma) \, d\epsilon_\gamma
\end{eqnarray}
is the overall emission probability for an electron with the Lorentz factor $\gamma'$,
$\epsilon_\ell = \hbar \omega / mc^2$, $\omega$ is the frequency used for normalization of time.

The Boltzmann equation~\eqref{Boltzmann} can be solved numerically as follows. In finite-difference method the
distribution function $f'(\gamma')$ is represented as a vector, and the right-hand-side of the
Eq.~\eqref{Boltzmann} is represented as the product of a matrix and a vector. Then Euler method can be
used, and the computation of $f'(t', \gamma')$ from $f'(t' = 0, \gamma')$ is reduced to a matrix
exponentiation, that can be done with square-and-multiply algorithm that have logarithmic complexity
on the number of time steps. Then the distribution function in the initial reference frame can be
found from $f'(t', \gamma')$ with Lorentz transformation. For that one should neglect the
electron displacement in $K'$ (i.e., $x'(t') - x'(0)$) and assume that in $K'$ the angles $\varphi'$
between $x'$ axis and ${\mathbf v}_\perp'$ are uniformly distributed on the interval $[0, 2 \pi)$:
\begin{eqnarray}
    t = t' \Gamma, \\
    \label{gamma_from_gamma'}
    \gamma = \gamma' \Gamma (1 + V v_\perp \cos \varphi'),
\end{eqnarray}
\begin{multline}
    f(\gamma) \propto \int f'(\gamma') \frac{d \gamma'}{d \gamma} \, d\varphi' \\
    = \int \frac{f'(\gamma')}{(1 + V v_\perp \cos \varphi')} \, d\varphi',
\end{multline}
where the integration should be performed over the path determined by the value of $\gamma$ and
Eq.~\eqref{gamma_from_gamma'}; $\Gamma = (1 - V^2)^{-1/2}$. It is worth noting that for correctness
of the method the step along $\gamma'$ in the finite-difference scheme should be much smaller than the
width of the emission spectrum. Thus, especially small step of $\gamma'$ should by used in
the classical regime.

Figure~\ref{cornu} (b) demonstrates the electron spectra
in the crossed fields
Eq.~\eqref{crossed_fields1}-\eqref{crossed_fields2}
with $a_0 = 50.3$ and $\lambda = 1.1 \text{
 nm}$ used for the normalization. The electrons initially (at $t = 0$) move along $x$ and $z$ axes
($v_x \simeq 0.8$, $v_z \simeq 0.6$) and have
Lorentz factor $\gamma_0 = 63$. Curves A and C are obtained by Eq.~\eqref{Boltzmann} for $t = 22$
and $t = 49.5$, respectively. Curves B and D represent the spectra of $8000$ particles whose
trajectory is computed by Vay's pusher coupled with Monte Carlo event generator, for $t = 22$ and
$t = 49.5$, respectively.

In $K'$ the parameters of the simulations yield the quantum parameter $\chi'(t' = 0) = 2$, and if
Landau--Lifshitz radiation reaction is used, $\chi'$ drops down to $\chi'(t = 22) = 0.4$ and
$\chi'(t = 49.5) = 0.1$ (see Eq.~\eqref{cornu_g}). However, initially $\chi' \gtrsim 1$ that leads
to wide emission spectrum and wide resulting spectrum of the electrons. Moreover, the overall emission
probability is not very high and a significant fraction of electrons do not emit photons at all.
These electron fractions form peaks clearly seen on the curves A and B. The position of the peak on
the curve A corresponds to non-emitting electrons with $v_x' = -0.5$ that according to
Eq.~\eqref{gamma_from_gamma'} gives $\gamma \approx 38$. However, in Monte Carlo simulation at $t =
22$ the distribution of electrons over $\varphi'$ is far from the uniform one, and most of the
non-emitting electrons moves with $v_x \approx 0.5$ leading to the peak at $\gamma = \gamma(t =
0)$. Thus, the difference of curves A and B comes from the assumption of uniform electron
distribution over the angle $\varphi'$. This assumption becomes more reliable at later times ($t =
49.5$), and the difference between two methods of the spectra computation vanishes (see curves C
and D).

Therefore, the results of the Vay's pusher coupled with the Landau--Lifshitz radiation reaction force
or with the Monte Carlo event generator (that uses some approximate expression for fast computation of
the emission probability) coincides well with the results obtained by other methods.

\section{Multipole wave}
\label{appendix-multipole-wave}

In the cylindrical coordinates the vector potential $\mathbf A$ of the current loop obeys the following
equation:
\begin{equation}
    \Delta A_\varphi - \partial_t^2 A_\varphi = -j_\varphi,
\end{equation}
where we assume that the $z$-axis is the axis of the loop, thus $A_r = A_z = 0$. The solution of
this equation for the harmonic current $j_\varphi \propto \exp(-i \omega t)$ (obviously, in the
normalized units $\omega = 1$) can be found using Green's function as follows~\cite{Jackson}:
\begin{equation}
    \label{A_varphi}
    A_\varphi = -\frac{I_0 r_\ell}{\pi \bar z} \int_0^{\pi/2} \frac{\cos(2 \psi)}{s} \exp(-i t + i
    s \bar z) \, d\psi,
\end{equation}
where $I_0$ is the current amplitude, $\bar z = [z^2 + (r + r_\ell)^2]^{1/2}$, $s = (1 - \kappa
\sin^2 \psi)^{1/2}$ and $\kappa = 4 r r_\ell / {\bar z}^2$. Then the electric and magnetic fields
can be found from the Eq.~\eqref{A_varphi}.

To obtain the field of a multipole wave that is fully absorbed by the current loop, the
substitution $t \to -t$, $B \to -B$ is made. Then the fields are computed on the $r-z$ lattice, and their
values are used for the interpolation in the numerical solution of the equations of the electron
motion.

\bibliography{main}

\end{document}